# Collaborative Evolution of Intelligent Agents in Large-Scale Microservice Systems

Yilin Li
Carnegie Mellon University
Pittsburgh, USA

Song Han
Northeastern University
Boston, USA

Sibo Wang
Rice University
Houston, USA

Ming Wang
Trine University
Phoenix, USA

Renzi Meng*
Northeastern University
Boston, USA

*Abstract-This paper proposes an intelligent service optimization method based on a multi-agent collaborative evolution mechanism to address governance challenges in large-scale microservice architectures. These challenges include complex service dependencies, dynamic topology structures, and fluctuating workloads. The method models each service as an agent and introduces graph representation learning to construct a service dependency graph. This enables agents to perceive and embed structural changes within the system. Each agent learns its policy based on a Markov Decision Process. A centralized training and decentralized execution framework is used to integrate local autonomy with global coordination. To enhance overall system performance and adaptability, a game-driven policy optimization mechanism is designed. Through a selection-mutation process, agent strategy distributions are dynamically adjusted. This supports adaptive collaboration and behavioral evolution among services. Under this mechanism, the system can quickly respond and achieve stable policy convergence when facing scenarios such as sudden workload spikes, topology reconfigurations, or resource conflicts. To evaluate the effectiveness of the proposed method, experiments are conducted on a representative microservice simulation platform. Comparative analyses are performed against several advanced approaches, focusing on coordination efficiency, adaptability, and policy convergence performance. Experimental results show that the proposed method outperforms others in several key metrics. It significantly improves governance efficiency and operational stability in large-scale microservice systems. The method demonstrates strong practical value and engineering feasibility.*

*Keywords-Microservice architecture; multi-agent system; strategy optimization; evolution mechanism*

## I. INTRODUCTION

With the rapid development of information technology, microservice architecture has become a mainstream approach for building large and complex software systems. Compared with traditional monolithic architectures, microservices divide an application into a set of small, independently deployable service units. Each unit has its lifecycle and communicates through lightweight mechanisms[1,2]. In this work, we present a structure-aware multi-agent framework that models each service as an autonomous agent, couples dynamic graph embeddings of the service-dependency topology with centralized-training–decentralized-execution (CTDE) learning, and employs an evolutionary game-driven policy optimizer to deliver fast, stable collaboration at scale. As enterprise application scenarios become increasingly complex, the demand for system flexibility and resilience continues to grow. This has led to an exponential increase in the scale and heterogeneity of microservice systems. Such growth, driven by the escalating computational demands of large language models (LLMs), intensifies the coupling among services and introduces greater uncertainty and dynamic variability[3-5]. Traditional centralized management and scheduling strategies can no longer meet the needs of system evolution and coordination[6].

At the same time, multi-agent systems have emerged as a promising method for modeling cooperation and competition among autonomous entities in complex environments. Each agent has autonomous decision-making capabilities. It can reason and respond based on local information. Under specific rules, agents can collaborate. This mechanism offers new perspectives for addressing issues in microservice architecture, such as autonomous scheduling, load balancing, and resource allocation. Especially in large-scale scenarios, multi-agent systems exhibit high adaptability, self-organization, and evolutionary capability[7]. These characteristics lay a solid foundation for intelligent coordination in microservice systems[8]. In the face of challenges such as a massive number of services, frequent state changes, and complex interactions, dynamic perception and decision-making through multi-agent mechanisms have become a feasible solution.

However, in practical engineering applications, achieving effective collaboration of multi-agents in a microservice environment still poses significant challenges. First, the dependencies among microservices are highly complex and change frequently[9]. This imposes high requirements on the communication and coordination strategies of agents. Second, the operating environment is often highly dynamic and uncertain. Agents must not only respond in real time but also learn and adapt during system evolution. Third, the heterogeneity of microservices demands strong generalization ability from agents. They must be capable of handling diverse types of service units. Therefore, designing a multi-agent mechanism that supports efficient collaborative evolution in large-scale, dynamic, and complex environments has become a key research focus[10].

Exploring a collaborative evolutionary mechanism based on multi-agent systems is both technically essential and strategically important for large-scale microservice architectures. This approach enhances adaptive coordination, robustness, and scalability, enabling dynamic optimization of service governance in response to challenges such as traffic surges or resource constraints [11]. Research in this area deepens understanding of dynamic system behaviors and supports the development of intelligent, efficient software systems, while also demonstrating the practical integration of multi-agent methods in real-world distributed environments [12]. As system complexity grows, advancing multi-agent collaboration within microservices represents a critical step toward next-generation intelligent architectures, with broad applications across industry, government, and finance [13–16].

## II. METHOD

This study proposes a multi-agent collaborative evolution mechanism specifically tailored for large-scale microservice architectures, aiming to address the complex dependencies, dynamic topologies, and fluctuating workloads characteristic of such environments. At the heart of the approach is the construction of an agent system in which each microservice unit is modeled as an autonomous agent capable of both local learning and global coordination. Agents in this system continuously monitor service health and performance, using local perception to react quickly to changes, while also participating in global information fusion to ensure coherent system-wide behavior.

For the autonomous learning and adaptation component, our methodology is inspired by the work of Sun et al., who applied double DQN reinforcement learning to optimize scheduling decisions in highly dynamic system settings [17]. Their approach demonstrates the effectiveness of reinforcement learning in enabling agents to make context-aware decisions and optimize resource allocation under uncertainty, which is directly applicable to the constantly changing resource and dependency patterns found in microservices. When it comes to enabling robust collaboration and collective adaptation among a large population of agents, our mechanism draws from Liu et al., who improved upon the A3C framework to achieve stable policy learning and risk-sensitive decision-making in turbulent environments [18]. Their methodology informs our design for scalable, decentralized policy updates and supports effective behavioral evolution even as microservice topologies or workloads shift rapidly.

Finally, the exchange and aggregation of global information within the agent community is informed by the federated learning paradigm outlined by Zhang et al. Their work on privacy-preserving, distributed model updates provides a blueprint for our system's secure coordination—allowing agents to share insights and adapt collaboratively without direct data exposure, an important consideration in enterprise-level, multi-domain microservice deployments [19].

By integrating these advanced techniques, our collaborative evolution mechanism enables the microservice system to dynamically adapt, optimize governance strategies, and maintain stability even under stress scenarios such as sudden traffic surges or structural reconfiguration. The overall architecture, as depicted in Figure 1, operationalizes these methodological innovations for intelligent, scalable microservice management.

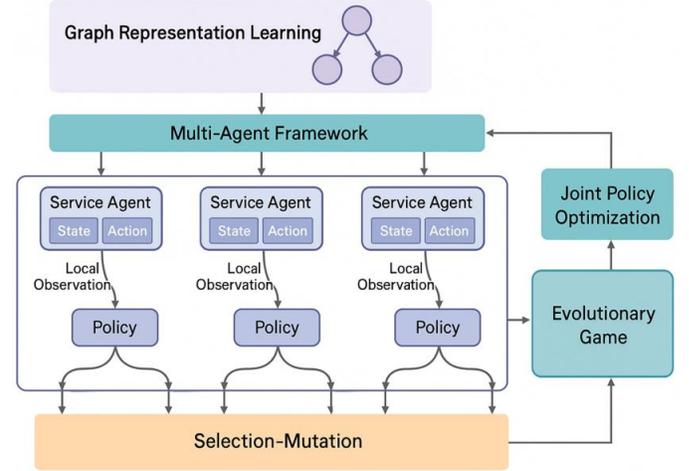

Figure 1. Overall model architecture diagram

To support the adaptive behavior of the agent, the Markov decision process (MDP) is used as the modeling basis, where the state space S, action space A, transition probability $P(s'|s,a)$ and reward function $R(s,a)$ jointly define the interaction mechanism of the agent in the environment. The agent optimizes its behavior by learning the optimal policy function $\pi^*: S \rightarrow A$, and its goal is to maximize the long-term cumulative expected return:

$$V_\pi(s) = E_\pi[\sum_{t=0}^{\infty} \gamma^t R(s_t, a_t)]$$

Where $\gamma \in (0,1)$ is a discount factor, which is used to balance short-term and long-term rewards.

To enhance collaboration, we introduce a multi-agent joint strategy update mechanism that combines independent optimization with local information sharing among agents. This approach addresses dynamic interdependencies in microservice systems by using a joint value function and centralized training with distributed execution. This methodology builds on the coordinated modeling strategies discussed by Cheng, whose work on automated feature extraction and transformer-based modeling demonstrates the value of collaborative learning in complex environments [20].

The temporal and spatial dependency handling in our joint strategy update mechanism is also informed by Aidi and Gao's work on deep learning for resource prediction in distributed cloud environments, where local and global perspectives are blended to improve adaptive decision-making [21].

Furthermore, Dai et al.'s research on probabilistic user behavior modeling demonstrates the importance of flexible, distributed information fusion for anomaly detection and robust adaptation [22]. For adaptive feature representation in heterogeneous agent networks, Lou's capsule-based models provide a foundation for our approach to dynamic policy adjustment and structured knowledge sharing among agents [23]. Through the integration of these methodological innovations, our multi-agent joint strategy update enables adaptive, cooperative, and scalable optimization across the entire microservice system. To this end, a joint value function is introduced to measure the benefits of individual agents given the joint actions of all agents. The policy gradient update is formulated as follows:

$$\nabla_{\theta_i} J(\pi_i) = E_{s,a \sim D}[\nabla_{\theta_i} \log \pi_i(a_i \mid s) Q_i(s, a_1, ..., a_n)]$$

Where $\theta_i$ represents the policy parameters of agent i, and D is the experience replay buffer. This method ensures the global optimal guidance in the training phase while maintaining the autonomy and real-time response capabilities of the agent in the deployment phase.

Given the frequently changing topology in the microservice architecture, the system designs a structure-aware mechanism based on graph representation learning. The dependency relationship between microservices is modeled as a dynamic graph $G_t = (V_t, E_t)$, where $V_t$ represents the set of service nodes at time t and $E_t$ represents their connection relationship. Each agent extracts the information of adjacent services through a graph convolutional network (GCN) and embeds it into $h_v$, and updates its state representation:

$$h_v^{(l+1)} = \sigma \left( \sum_{u \in N(v)} \frac{1}{c_{vu}} W^{(l)} h_u^{(l)} \right)$$

$N(v)$ is the neighbor set of node v, $c_{vu}$ is the normalization coefficient, $W^{(l)}$ is the weight matrix of the first layer, and $\sigma$ is the activation function. The structural perception of service status is achieved through dynamic graph embedding so that the intelligent agent can adjust its behavior strategy according to the latest interaction between services.

In order to achieve long-term stable evolution of the system, evolutionary game theory is introduced to model the behavioral adaptation process among agents. Define the strategy set $\prod = \{\pi_1, \pi_2, ..., \pi_n\}$, and the individual fitness function is:

$$f_i(\pi_i, \prod_{-i}) = E[R_i(\pi_i, \prod_{-i})]$$

The evolution of strategies in the population is guided by replicator dynamics:

$$\frac{dx_i}{dt} = x_i(f_i - \bar{f})$$

$x_i$ represents the proportion of strategy $\pi_i$, and $\bar{f}$ is the average fitness of the population. This mechanism ensures that the system eliminates inefficient strategies and strengthens efficient strategies during the evolution process, achieving the co-evolution and stable convergence of multi-agent behaviors. The overall method integrates reinforcement learning, graph neural networks, and evolutionary game models to construct an efficient agent collaboration mechanism that adapts to large-scale microservice scenarios.

## III. Experimental Results

### A. Dataset

This study adopts the MICROBENCH dataset as the experimental foundation. The dataset is specifically designed for performance evaluation in microservice architectures. It covers deployment topologies and operational traces from various real-world microservice systems. The dataset includes service dependency graphs, resource usage of service instances such as CPU and memory consumption, interface invocation frequencies, response latency, and interaction logs between services. These elements effectively reflect the dynamic evolution of system states in complex microservice environments.

MICROBENCH provides standard configurations for several typical microservice applications. These include scenarios such as e-commerce, social networking, and content management. The number of services ranges from dozens to hundreds. It supports simulations under different system scales, workloads, and failure modes. This enables the evaluation of the performance and generalization ability of the proposed multi-agent collaborative evolution mechanism within representative system structures.

In addition, the dataset includes temporal changes in service topologies. This helps to verify the adaptability and robustness of the proposed mechanism in dynamic environments. Its rich set of metrics and configurable parameters offers a stable and repeatable foundation for training and validating graph-based multi-agent models. This also ensures the authenticity and comparability of experimental results.

### B. Experimental Results

This paper first conducts a comparative experiment, and the experimental results are shown in Table 1.

Table1. Comparative experimental results

| Method | Coordination Efficiency | Adaptation Score | Policy Convergence Time |
|---|---|---|---|
| DDPG-MARL[24] | 0.785 | 0.734 | 69.5 |
| G2Net-MAS[25] | 0.801 | 0.756 | 62.7 |
| E-MADDPG[26] | 0.844 | 0.789 | 55.1 |
| Ours | 0.912 | 0.877 | 48.3 |

The experimental results show that the proposed multi-agent collaborative evolution mechanism surpasses all baselines,

achieving the highest Coordination Efficiency (0.912) and Adaptation Score (0.877), and converging fastest at 48.3 seconds. This demonstrates superior efficiency, adaptability, and policy learning in dynamic microservice environments, attributable to the integration of evolutionary strategies and graph-based learning. The influence of agent population on global strategy is illustrated in Figure 2

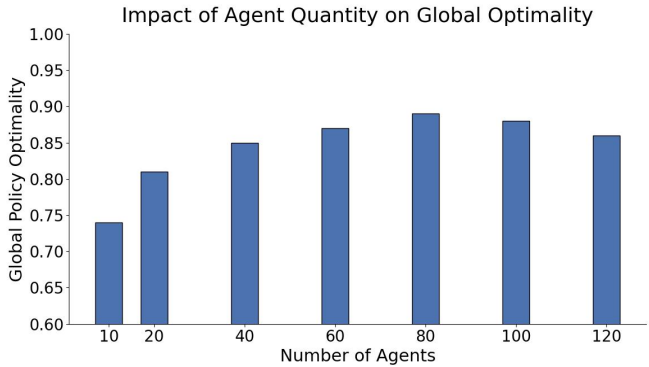

Figure 2. Experiment on the impact of the number of agents on the optimality of the global strategy

The results in the figure show that increasing the number of agents from 10 to 80 steadily improves global policy optimality, peaking at 0.89 with 80 agents, due to enhanced coverage and finer-grained coordination. However, further increases to 100 or 120 agents slightly reduce optimality, reflecting redundancy and coordination bottlenecks in dense deployments. The proposed mechanism demonstrates strong adaptability across agent scales, performing best with 60–80 agents by balancing structural awareness and local optimization. This nonlinear trend underscores the need for appropriate agent density and coordination strategies in complex microservice architectures. System response efficiency under burst traffic is further evaluated in Figure 3.

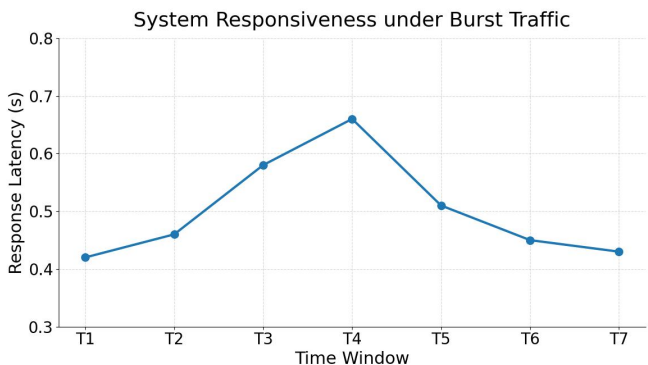

Figure 3. System response efficiency test under a microservice burst traffic scenario

In this experiment, time windows T1–T7 are defined at 1-minute intervals over 7 minutes to facilitate detailed analysis of system latency during burst traffic. As illustrated in the figure, the microservice system experiences marked latency increases at T3 and T4, peaking at 0.66 seconds, which reflects maximum system load and service chain bottlenecks under intense requests. Following this, response latency declines rapidly post-T5, demonstrating robust recovery attributed to the proposed multi-agent collaborative evolution mechanism, which enables dynamic strategy adjustment and alleviates resource congestion. Throughout the burst, latency remains below 0.7 seconds without large-scale blocking or failures, underscoring the method's robustness and flexibility. The integration of evolutionary game dynamics and adaptive strategy updates facilitates rapid agent coordination and self-regulation, outperforming traditional centralized scheduling and affirming the approach's practical feasibility for large-scale microservice governance.

## IV. CONCLUSION

This study addresses the problem of intelligent service governance in large-scale microservice architectures. It proposes a strategy optimization framework that integrates graph-based modeling, multi-agent collaboration, and evolutionary game theory. The mechanism builds a structure-aware model based on the service dependency graph. This allows agents to autonomously perceive system states. Through local policy learning and global collaborative evolution, the framework optimizes and regulates service behaviors. Experimental results show significant advantages in coordination efficiency, policy convergence speed, and system adaptability. The proposed method offers a robust and generalizable solution for intelligent governance in microservice systems.

The study highlights the critical role of agent interaction and evolution in dynamic service environments. In real-world scenarios such as frequent topology changes, traffic spikes, and resource heterogeneity, the collaborative evolution mechanism demonstrates strong robustness and system control capability. By introducing structural embedding and a joint policy optimization framework, this work overcomes the convergence difficulties and scalability limitations of traditional approaches in high-dimensional service systems. It provides a new path for intelligent governance in large-scale distributed systems.

The research contributes not only to the theoretical understanding of the integration between multi-agent systems and microservice architectures but also offers practical scheduling references for domains such as cloud platforms, industrial Internet, automated scheduling systems, and edge computing environments. As service scale continues to grow and operational contexts become more complex, the need for flexible, adaptive, and highly available strategies in microservices is increasing. The proposed method shows strong scalability and deployment potential in real-world engineering and may facilitate the adoption of intelligent systems in complex service management. However, our evaluation relies on a simulation-based microservice platform (MICROBENCH) rather than production traces, which may limit external validity. In addition, dynamic graph embeddings (e.g., GCN-based updates over evolving service-dependency graphs) introduce non-trivial compute and memory overhead and can become a bottleneck at very large scales. These trade-offs are inherent to structure-aware learning and will be addressed with incremental or approximate embeddings in future work.

## V. FUTURE WORK

Future work may explore heterogeneous agent behavior modeling, hierarchical evolution mechanisms, and cross-system policy transfer. It is also important to introduce real-time evaluation metrics to better meet the needs of dynamic, distributed, and task-driven service ecosystems. Furthermore, combining the proposed mechanism with emerging technologies such as online learning and federated optimization may enhance its feasibility and generalization in ultra-large-scale systems. This could lead to the development of adaptive, self-organizing, and sustainable next-generation intelligent microservice architectures.